# Large magnetocaloric effect and magnetoresistance behavior in $Gd_4Co_3$

Niharika Mohapatra, Kartik K Iyer and E.V. Sampathkumaran[*]

Tata Institute of Fundamental Research, Homi Bhabha Road, Colaba, Mumbai 400005, India.

Abstract

We report a large entropy change ($\Delta S$) below 300 K, peaking near $T_C$= 220 K, due to isothermal change of magnetic field, for $Gd_4Co_3$, with a refrigeration capacity higher than that for, say, $LaFe_{11.4}Si_{1.6}$, ordering magnetically in the same temperature range. A noteworthy finding is that the isothermal magnetization is nonhysteretic - an important criterion for magnetic refrigeration without loss. $\Delta S$ behavior is also compared with that of magnetoresistance.

## 1. Introduction

The interest on the topic of 'magnetocaloric effect (*MCE*)' – an effect arising in solids as a result of entropy variation due to the coupling of the spin system with the magnetic field - has picked up considerable momentum in recent years due to various advantages in using materials with large *MCE* in refrigeration. A good number of such materials with large *MCE* in different temperature ranges have been reported in the literature [1] at the magnetic ordering temperatures. For many practical applications, one needs to have materials with significant effect near room temperature and hence the materials with correspondingly high magnetic ordering temperatures are preferred. In addition, it is desirable [2, 3] that *MCE* is large over a wide temperature range, rather than at the magnetic transition only, for applications and also without any hysteresis loss. With these in mind, we have carried out magnetic measurements on the compound, $Gd_4Co_3$, crystallizing in the $Ho_4Co_3$-type hexagonal structure (space group: $P6_3/m$) [4] and ordering [5-7] magnetically at ($T_C=$) 220 K. The main finding is that this compound exhibits large *MCE* characteristics desirable for applications, that too without any hysteresis in isothermal magnetization (*M*).

## 2. Experimental

The sample in the polycrystalline form was prepared by repeated melting of stoichiometric amounts of high purity (>99.9%) constituent elements in an arc furnace in argon atmosphere. The weight loss after final melting was less than 0.1%. The ingot was homogenized in an evacuated sealed quartz tube with the heat treatment as in Ref. 7 (600 C for 2 h, 635 C for 24 h and at 650 C for 48 h). The sample was characterized by x-ray diffraction (Cu $K_\alpha$) and found to be single phase ($a=$ 11.61 Å; $c=$ 4.048 Å). The M measurements were performed employing a vibrating sample magnetometer (Oxford Instruments, UK) in the temperature interval 2 – 300 K. We have also performed heat-capacity (C) measurements (2-270 K) employing a commercial (Quantum Design, USA) physical property measurements system. The same instrument was used to carry out electrical resistivity ($\rho$) studies (2 - 300 K) as a function of magnetic field (*H*).

## 3. Results and discussion

The results of magnetization measurements, obtained in the presence of two selected magnetic fields, as a function of temperature are shown in figure 1. We did not find any difference between the curves for the zero-field-cooled and field-cooled conditions of the specimen during measurements. The magnetic susceptibility ($\chi$) exhibits Curie Weiss behavior above 225 K and the effective moment obtained from the linear region (see figure 1, inset) in the plot of $1/\chi$ turns out to be 8.25 $\mu_B$/Gd. This value is larger than that expected for $Gd^{3+}$ ion (7.94 $\mu_B$), which is attributed to Co. The value of paramagnetic Curie temperature turns out to be 220 K. The magnetization exhibits an upturn below 220 K attributable to the onset of magnetic ordering of a ferromagnetic type. This upturn is sharp if measured in low fields, e.g., 100 Oe. In addition, there is another weak upturn at 163 K, the sharpness of which also depends upon the magnitude of the applied field (that is, sharper at low fields; compare the feature for *H*= 100 Oe shown in the inset with that for *H*= 5 kOe), attributable to another magnetic transition, confirming an earlier report [7]. We have also performed isothermal *M* measurements up to 100 kOe (Fig. 2) and we find that, at low temperatures, following initial sharp rise at low fields, there is a very weak nearly linear increase of *M* at higher fields, as though there is a tendency for saturation. The value of the saturation moment, ($M_s(0)$), obtained by linear extrapolation of the high field data to zero field, say at 2.5 K, turns out to be less by 0.5 $\mu_B$ per Gd ion compared to the free ion value for Gd. This implies that Co acquires a magnetic moment due to the molecular field of Gd ions and it couples antiferromagnetically with Gd. It therefore appears that the net magnetic structure is in fact ferromagnetic. These features are in agreement with those reported in the literature [7]. The finding we would like to stress is that the isothermal $M$ curves are non-hysteretic in the entire temperature range of investigation.

We now address MCE behavior of this compound. For this purpose, we measured isothermal $M$ at several temperatures in close intervals (typically 3 to 5 K), and we show these curves at few selected temperatures only in figure 2. The isothermal entropy change, $\Delta S$ [= $S(H_2)$-$S(H_1)$], for a change of the magnetic field from $H_1$ to $H_2$ is a measure of *MCE*, and this can be determined from the well-known Maxwell's relationship. The data thus derived are shown in figure 3 for a change of magnetic field from 0 to $H$. The sign of $\Delta S$ is negative, which means that heat is liberated when the magnetic field is changed adiabatically. It is clear that $\Delta S$ peaks at $T_C$ and the values are large. If one assumes a theoretical density of 8.5g/cc, we get an idea [1] about the values in volume units. Thus, for instance, for a variation $H$ from 0 to 50 kOe, at $T_C$, the value of $\Delta S$ turns out to be about 42 mJ/ccK. This value is comparable to that of many Gd based alloys in the temperature range 200 – 300 K (see figure 3 in Ref. 1). A notable observation is that the peak of $\Delta S$ versus $T$ is rather broad extending over a wide temperature range, say 100 to 250 K. Quantitatively, for an inference for a possible application for refrigeration, a parameter called 'refrigeration capacity (*RC*)', is defined in two different ways [8,9]. Here, for comparative purposes, we stick to the method of Ref. 8:

$$RC = \int_{T_{cold}}^{T_{hot}} \Delta S(T) \cdot dT$$

Here, $T_{cold}$ and $T_{hot}$ correspond to the two temperatures at which the $\Delta S$ value is half of the peak value. It is found that the value of *RC* is 4.3 J/cc for $H$= 0 to 50 kOe. For the sake of comparison, the corresponding value for Gd is 5.5J/cc [1,2]. If one wants to make a comparison with the data known in the literature for other materials ordering in the same temperature range as the present compound, the corresponding value for the compound, $LaFe_{11.7}Si_{1.6}$ (tabulated in Ref. 2, ordering magnetically around 208 K) is relatively lower, about 3.5 J/cc.

Further support to the above conclusions is obtained from the heat-capacity data measured in zero field as well as in 50 kOe (Figure 4). In zero field, anomalies in C(T) are observed around the magnetic transitions near 163 and 220 K. The values of $\Delta S$ obtained from these data are plotted in figure 3, and the curve almost overlaps with that from M data for H= 0-50 kOe. We also plot the adiabatic temperature change, $\Delta T$, defined as T(S,50kOe)-T(S,0), in the inset of figure 4, which shows qualitatively similar features as $\Delta S$ (We also note a weak peak around 240 K in C(T), the origin of which is not clear to us).

Recently, the existence of a relationship between $\Delta S$ and $H$ has been proposed theoretically [10, 11]. For instance, it has been derived on the basis of renormalization approach and established experimentally [11] that there is a well-defined relationship between $\Delta S$ and $H$, and for magnetic materials with second order phase transition, the experimental data obey the relation, $\Delta S$= - $kM_s(0)h^{2/3}$-$S(0,0)$, where $h$ is the reduced field (given by $\mu_0\mu_B H/k_B T_C$). $k$ is a constant and $M_s(0)$ is the saturation magnetization at low temperatures. The $\Delta S$ data for about a dozen materials were fitted to this equation successfully, but there is a finite value of $S(0,0)$ ranging from -0.2 to -1.06/kgK, the physical significance of which is not yet clear. In order to address this issue in the present material, we plot the peak value of $\Delta S$ (obtained from magnetization) as a function of $h^{2/3}$ in the inset of figure 3. It is found that the plot is linear, barring a small deviation at low $h$ values (as in Ref. 11). The values of $S(0,0)$ and the coefficient of $h^{2/3}$ term are found to be -1.2 and -106 J/kgK very close to the values reported in table 1 of Ref. 11. The value of $k$ is about 0.6. [For the sake of direct comparison with Ref.11, in the inset of figure 3, we present $\Delta S$ in the units of J/kgK].

We now present the results of our magnetoresistance (*MR*) measurements, primarily to compare with $\Delta S(T)$ behavior. As reported before [7], there is a well-defined drop at $T_C$ in the plot of electrical resistivity versus temperature, $\rho(T)$. Thus the feature is smeared out at high fields, as shown in figure 5 (top). To get a better idea about magnitude and sign of MR, we have also measured as a function of magnetic field at selected temperatures (see figure 5, bottom). While *MR* [defined as $\{\rho(H)- \rho(0)\}/\rho(0)$] is negative above about 50 K, we find that there is a sign-crossover of *MR* below 50 K. It is possible that positive values of *MR* below about 50 K is a signature of dominating antiferromagnetic component in zero field due to antiferromagnetic coupling between Gd and Co moments as indicated by $M_s(0)$. Such changes in magnetism, in addition to the transition at 163 K could account for the broadened $\Delta S(T)$ curve. There are in fact qualitative changes in the shapes of the plots (see the data below about 20 kOe) in figure 5 as the temperature is lowered across $T_C$; that is, at $T_C$, there is a sharper change for an initial application of $H$, whereas at lower temperatures, the variation of *MR* with $H$ is more gradual. The data curves of MR versus $H$ were fitted to the form, MR= $a-bH^n$ (where $a$ and $b$ are coefficients) and the value of $n$ is smaller than the expected value of 2 for the paramagnetic state, e,g., 1.4 at 300 K, presumably due to persistence of short-range magnetic correlations well above $T_C$. The value of $n$ becomes 0.5 at $T_C$ (220 K) and it increases to about 0.7 at lower temperatures; the continuous lines through the data points in figure 5 are obtained by this fitting. In addition, we noted that the magnitude of *MR* keeps increasing with decreasing $T$ below 300 K for a given field (inferred also by comparing $\rho(T)$ plots of zero field and of 50 kOe shown in top portion of figure 5), peaking near $T_C$, tracking $\Delta S(T)$ behavior, as emphasized earlier [12]. This implies that there is a close relationship between MCE and MR [12, 13]. It is also interesting to note that the magnetoresistive response is rather large even at 300 K (e.g., about 2% for $H$= 120 kOe) though $T_C$ is much lower, similar to other unusual systems reported by us [14].

**4.     Conclusion**

We have identified a Gd-based intermetallic compound, $Gd_4Co_3$, with a large magnetic refrigeration capacity over a wide temperature range around 220 K, without any noticeable hysteresis loss. These characteristics suggest that this compound can also be explored for magnetic refrigeration in the relevant temperature range, particularly considering that this material is stable in air and easy to synthesize. The relationship between MCE and magnetic field is viewed in light of a recent theory in Ref. 11. The results also establish the existence of a close relationship between MCE  and MR.

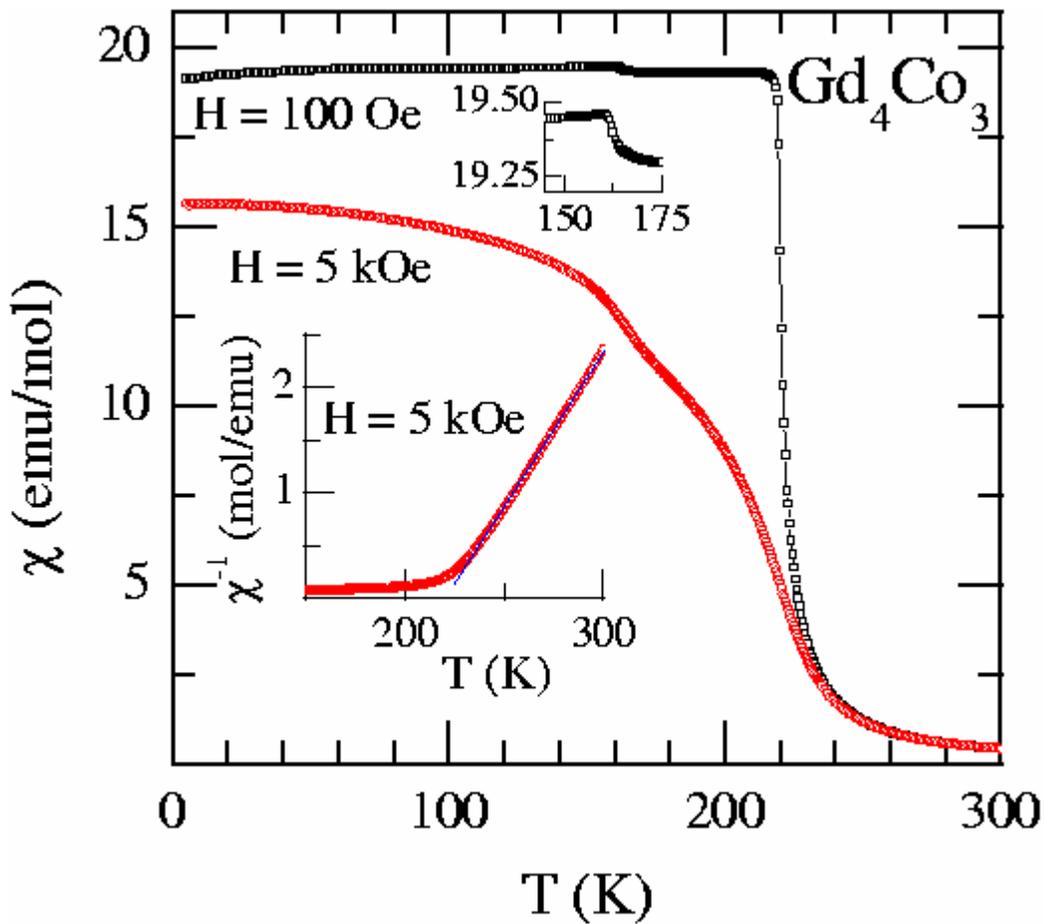

**Figure 1:**
(color online) (Top) Dc magnetic susceptibility as a function of temperature in the presence of 100 Oe and 5 kOe for $Gd_4Co_3$. There is a jump at 163 K for $H=$ 100 Oe due to a magnetic transition which is shown in an expanded form in the top inset. The bottom inset shows inverse susceptibility as a function of temperature and a straight line is drawn to show Curie-Weiss region.

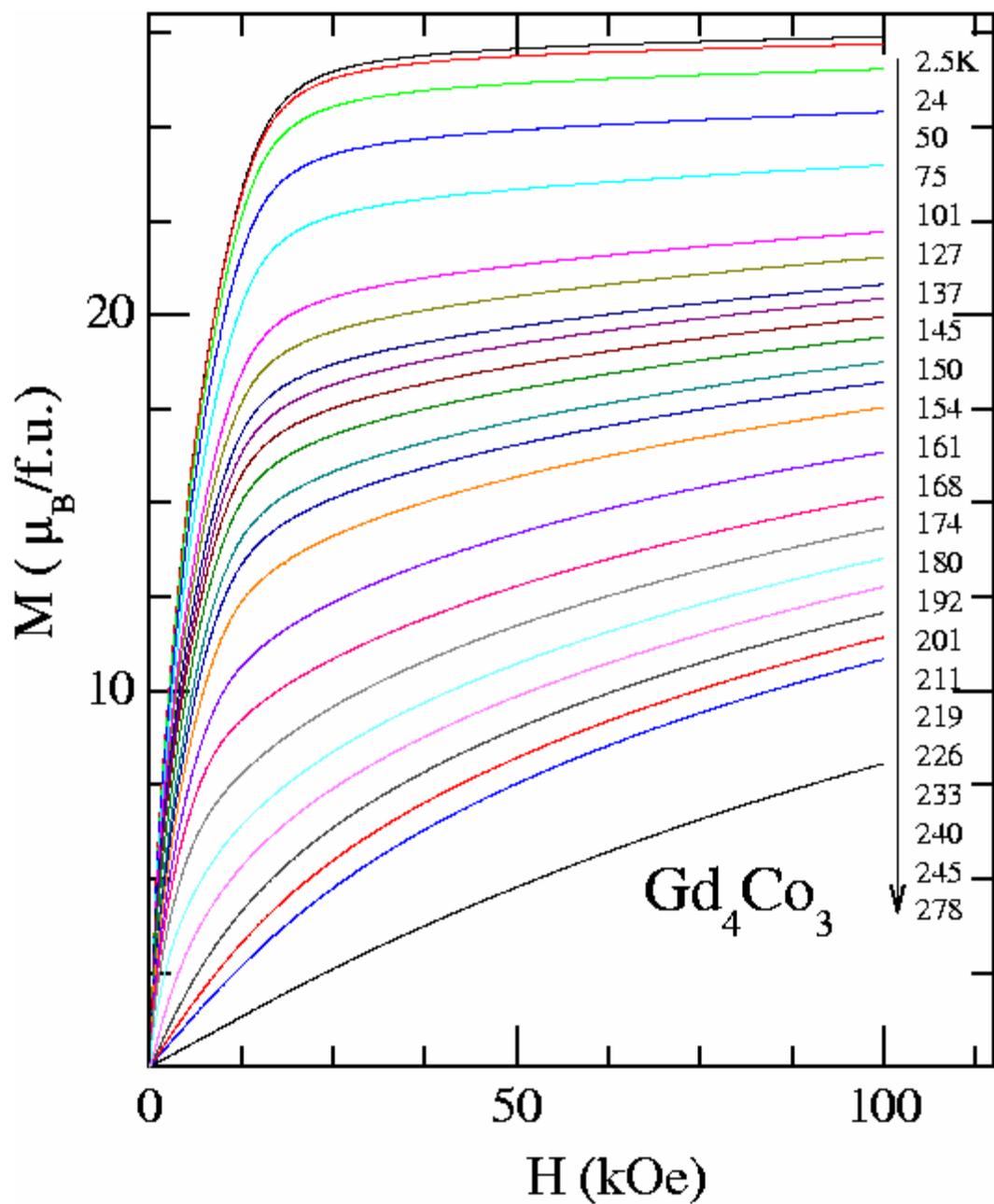

**Figure 2:**
(color online) Isothermal magnetization as a function of magnetic field at several temperatures for Gd$_4$Co$_3$. The curves for up and down field-cycles overlap for each temperature.

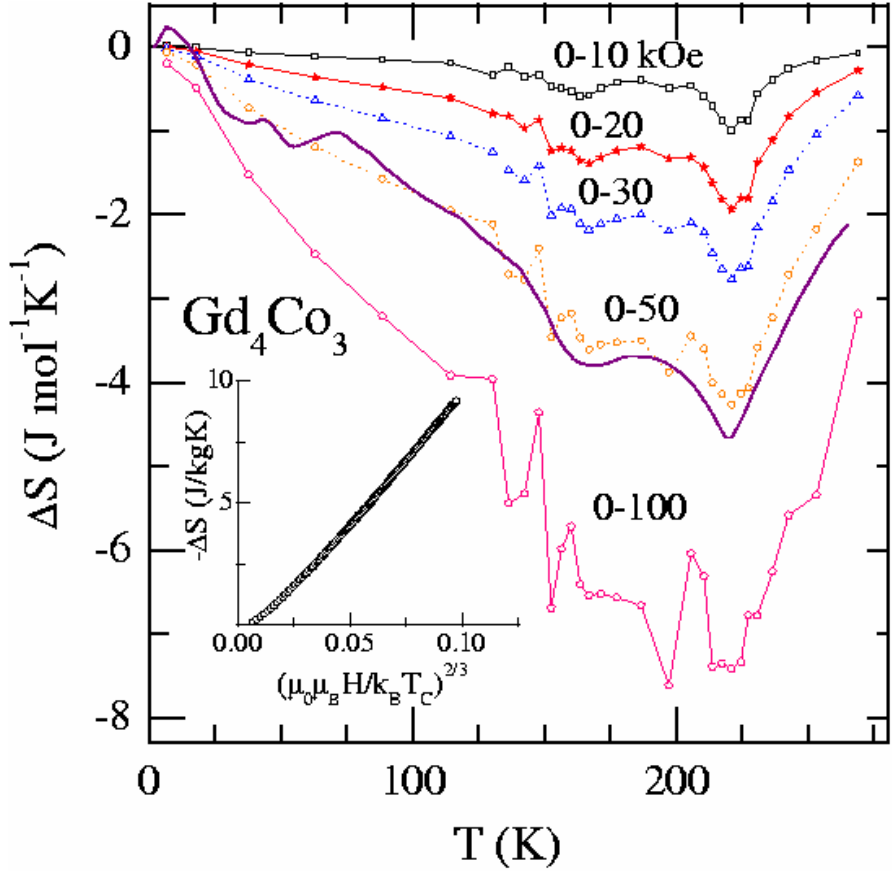

**Figure 3:**
(color online) The entropy change as a function of temperature for a variation of the magnetic field, obtained from the magnetization data employing Maxwell's equation, for $Gd_4Co_3$. The lines through the data points serve as guides to the eyes. Corresponding data obtained from heat-capacity (C) (see figure 4) are also plotted. The inset shows the existence of a relationship between the entropy change and the magnetic field as discussed in Ref. 11.

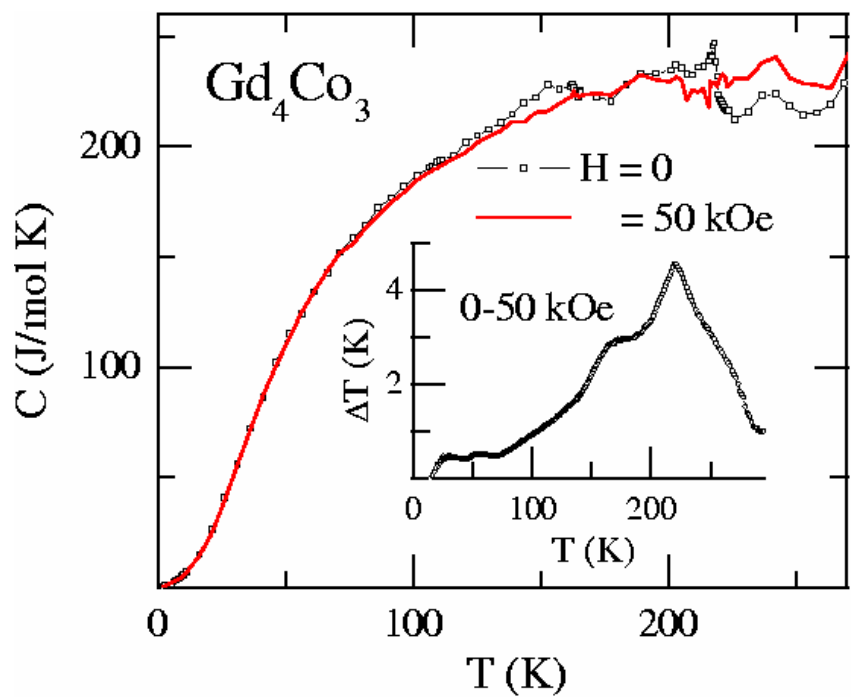

Figure 4:

(color online) Heat capacity as a function of temperature in zero field and in 50 kOe for $Gd_4Co_3$. Inset shows adiabatic temperature change derived from this data.

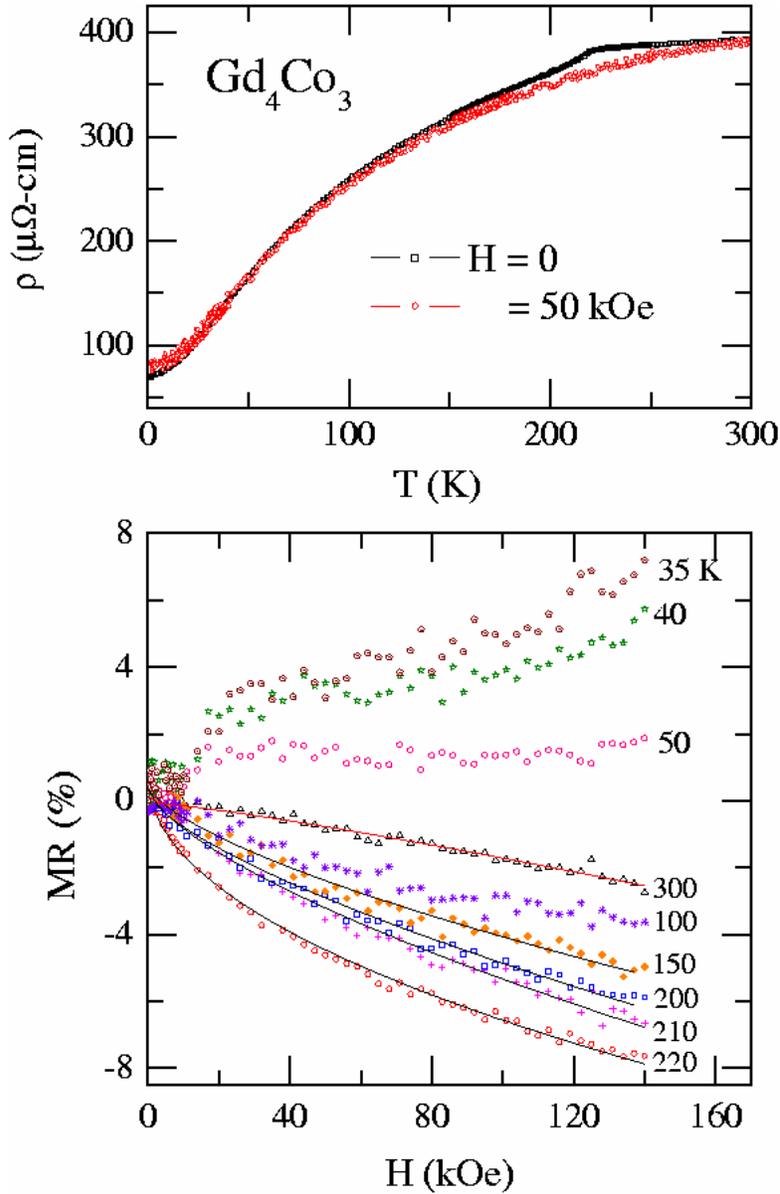

**Figure 5:**
(color online) (Top) Electrical resistivity as a function of temperature in the absence and in the presence of a magnetic field of 50 kOe. **(bottom)** Magnetoresistance as a function of magnetic field at various temperatures for $Gd_4Co_3$. The continuous lines through the data points are the fitted lines as described in the text.